\documentstyle[12pt]{article} 
\newcommand{\beqn}{\begin{equation}}
\newcommand{\eeqn}{\end{equation}}
\newcommand{\bea}{\begin{eqnarray}}
\newcommand{\eea}{\end{eqnarray}}

\begin{document}
\setcounter{page}{0}
\thispagestyle{empty}

\vskip 35pt

\begin{center}
{\LARGE\bf MASS BOUNDS FOR TRIPLET SCALARS OF THE LEFT-RIGHT
SYMMETRIC MODEL AND THEIR FUTURE DETECTION PROSPECTS}
\vskip 15pt
{\sf Anindya Datta$^{\heartsuit}$ and 
Amitava Raychaudhuri}$^{\dagger}$

\vskip 10pt

{\footnotesize Department of Physics, University of Calcutta, 
\\92 Acharya Prafulla Chandra Road, Calcutta 700009, India.}

\vskip 5pt
{\large\bf ABSTRACT} 
\end{center}

\sl 

The standard formulation of the Left-Right symmetric model involves
scalars transforming as a triplet under $SU(2)_L$. This multiplet
contains particles which are uncharged, singly-charged, and
doubly-charged. We derive a bound on the uncharged scalar mass of 55.4
GeV using results from LEP-II and find that a range upto 110 GeV may
be explored at the NLC at the 5$\sigma$ level.  We also discuss search
strategies for the singly- and doubly-charged scalars at the Tevatron
and the LHC.  Possible Standard Model backgrounds for the relevant
modes are estimated and compared with the signal. At the LHC, the
prospects of detecting the doubly-charged scalar are bright up to a
mass of 850 GeV while the 5$\sigma$ discovery limit of the
singly-charged mode extends to 240 GeV for an integrated luminosity of
100 fb$^{-1}$. At the Tevatron, with an integrated luminosity of 25
fb$^{-1}$, the doubly-charged state can be detected if its mass is
less than 275 GeV while the reach for the singly charged scalar is 140
GeV.

\vskip 40pt


\vskip 30pt

{\footnotesize e-mail: $^\heartsuit$anindya@mri.ernet.in,
$^ \dagger$amitava@cubmb.ernet.in}
\newpage
\rm 
\section{INTRODUCTION}
The quest for the higgs boson is one of the urgent missions of the
on-going and future particle physics experiments.  It is the key
missing ingredient of the Standard Model (SM) and is responsible
for the spontaneous breaking of the $SU(2)_L \times U(1)_Y$
symmetry of electroweak interactions. So far, direct experimental
searches for this scalar have proved fruitless and the absence of
a higgs signal at LEP-I and II yield a lower limit, 87.9 GeV
\cite{HigBound}, on its mass.

Though the SM has met with spectacular success under experimental
scrutiny, nonetheless, extensions beyond its gauge, scalar, or fermion
sectors have received much attention. In particular, the Left-Right
symmetric model (LRM), based on the gauge group $SU(2)_L \times
SU(2)_R \times U(1)_{B-L}$, has been put forward \cite{LRM} as an
equally viable alternative to the electroweak model with the added
virtue that, unlike in the SM, the non-conservation of parity is a
consequence of spontaneous symmetry breaking and not put in by hand.
In the minimal version of this model, the spectrum is enriched through
the introduction of right-handed partners of the observed gauge bosons
and neutrinos. These are significantly heavy, typically of the scale
$v_R \stackrel{>}{\sim}$ 1 TeV at which the LR symmetry is broken
spontaneously. Detailed phenomenology of the LRM has been extensively
examined and many constraints have been derived which restrict the
character of this model \cite{LRMP}.  In this work our attention will
revolve, in the main, around the scalar sector of the model which
consists of one bi-doublet, $\phi (1/2, 1/2,0)$, one right-handed
triplet, $\Delta_R (0,1,2)$, and one left-handed triplet, $\Delta_L
(1,0,2)$.  In the above, the $SU(2)_L$, $SU(2)_R$ and $U(1)_{B-L}$
quantum numbers of the fields are indicated in the parantheses.
$\Delta_R$ breaks the $SU(2)_R$ symmetry and can also generate a
majorana mass of the right-handed neutrino as required for the
``see-saw'' mechanism.  Out of the 20 degrees of freedom, after
spontaneous symmetry breaking to $U(1)_{EM}$, there remain 6 physical
neutral scalars (4 CP-even and 2 CP-odd), 2 physical charged scalars
(4 degrees of freedom), and 2 doubly-charged scalars (4 degrees of
freedom).  Except one CP-even neutral state, the others have masses at
the $v_R$ scale if the couplings of the scalar potential bear no
relation to each other.  The former, originating from the bi-doublet,
is the analogue of the higgs boson of the SM and has similar coupling to
the gauge bosons, save the suppression by a mixing angle factor which
is typically very close to unity.  A possible relationship among the
couplings in the scalar potential discussed later, which can arise
naturally from some larger symmetry, can keep the masses of some of
the other scalars, specifically from the left triplet, in the scale of
$m_W$, in which case they may be produced in the present and upcoming
colliders. In this work we examine the mass limits for these particles
and the prospect of their detection at the Tevatron and the LHC.

The plan of the article is as follows. In section 2, we discuss
in detail the LRM lagrangian, couplings, and scalar mass matrices
relevant to our analysis.  In section 3, we begin with the pair
production of the neutral triplet scalar at LEP-II. The triplet
scalars can only have majorana-type coupling to the leptons and
the neutral member couples only to neutrinos and  decays invisibly.
We utilize its production in association with a photon to set a
bound on its mass using the measured LEP cross-section of the
photon plus missing energy channel. In Section 4, we
turn to hadron colliders and examine the feasibility of
observation of the singly-charged scalars, which decay dominantly
to leptons, along with an analysis of possible SM backgrounds at
the Tevatron and the LHC.  The production and detection of
doubly-charged scalars are discussed in Section 5.  The
conclusions are in Section 6.

\section{LAGRANGIAN AND THE RELEVANT  COUPLINGS}

We begin this section by reviewing the salient features of the
minimal version of the LRM with an emphasis on the scalar sector.
A convenient representation of the scalar fields is given in terms of 
$2\times 2$ matrices: \\ 

$ 
\phi \equiv \pmatrix{\phi_1 ^0 & \phi_1
^+ \cr \phi_2 ^- & \phi_2 ^0} $,   $
\Delta_L \equiv \pmatrix{\delta_L ^+ /\sqrt{2} & \delta_L ^{++} \cr
\delta_L ^0 & -\delta_L ^+ /\sqrt{2}} $,  $
\Delta_R \equiv \pmatrix{\delta_R ^+ /\sqrt{2} & \delta_R ^{++} \cr
\delta_R ^0 & -\delta_R ^+ /\sqrt{2}} 
$
\\

%A neutral field $\phi^0$ is written in terms of its real and imaginary
%componant as :

% $\phi ^0 = (\phi_0 ^r + i \phi_0 ^i)/ \sqrt{2}$.

Under $SU(2)_{L,R}$ gauge transformations:
\beqn
\phi  \rightarrow U_L \phi U_R ^\dagger, \;\;
\Delta_L  \rightarrow U_L \Delta_L U_L ^\dagger, \;\;
\Delta_R  \rightarrow U_R \Delta_R U_R ^\dagger 
\eeqn
where $U_{L,R}$ are the appropriate $2 \times 2$ unitary matrices. 

The gauge symmetry breaking proceeds in two stages. In the first
stage, $\delta_R ^0$, the electrically neutral component of
$\Delta_R$, acquires a vacuum expectation value ({\em vev})
$v_R$ breaking the gauge symmetry down to $SU(2)_L \times
U(1)_Y$. The masses of the $W_R$ and $Z^\prime$ gauge bosons and
that of the right-handed neutrino field are also driven by $v_R$.
$k$ and $k^\prime$, the $vev$s of the neutral members of
the bi-doublet serve the dual purpose of breaking the $SU(2)_L
\times U(1)_Y$ symmetry to $U(1)_{EM}$, thereby setting the mass
scale of the observed $W_L$ and $Z$ bosons, and of providing the
quark and lepton dirac masses.  $v_R$ is significantly larger
than $k,k^\prime$ so that right-handed gauge bosons are heavier
than the $W_L$ and $Z$. $\Delta_L$ is the LR symmetric
counterpart of $\Delta_R$. 
%The $vev$ of its neutral member,
%$v_L$, plays an important role in the $vev$ ``see-saw" relation,
%characteristic of the model (see below). 
$v_L$ must be much
smaller than $k,k^\prime$ in order that the deviation of $\rho$ (= 
$M_W^2/M_Z^2 \cos^2\theta_W$) from unity be very small,
as observed experimentally \cite{LEPWG}.
 
We now turn to the scalar potential. Under LR symmetry, $\phi
\leftrightarrow \phi^ \dagger ,\, \Delta_R \leftrightarrow
\Delta_L$ and also $\Psi_L \leftrightarrow \Psi_R$, where
$\Psi_{L,R}$ are the column vectors containing the left-handed
and right-handed fermionic fields of the theory.  Moreover, the
LR symmetry forbids any trilinear terms in the scalar potential.
Because of the non-zero values of the $B-L$ quantum numbers of
the triplet fields, they must appear in the quadratic
combination $\Delta_{i} \Delta_{j} ^\dagger$, where $i,j =
L,R$. The scalar potential satisfying these requirements can be
written as :
\beqn
{\cal V} = {\cal V}_\phi + {\cal V}_\Delta + {\cal V}_{\phi \Delta}
\eeqn
where,
\bea
{\cal V}_\phi  &=& - \mu_1 ^2 Tr( \phi^\dagger \phi)
     - \mu_2 ^2 [Tr( \tilde \phi \phi ^\dagger) + 
   Tr( \tilde \phi ^\dagger \phi)]
     + \lambda_1 [ Tr( \phi^\dagger \phi) ]^2 
     + \lambda_2 [ Tr( \phi^\dagger \tilde {\phi})^2 +  \nonumber \\
&&       Tr( \tilde {\phi}^\dagger {\phi}) ^2 ] 
     + \lambda_3 [ Tr( \phi^\dagger \tilde {\phi}) 
        Tr( \tilde {\phi}^\dagger {\phi}) ]
     + \lambda_4 Tr(\phi ^\dagger \phi)[Tr( \tilde \phi \phi ^\dagger) + 
       Tr( \tilde \phi ^\dagger \phi)] 
\eea
where $\tilde \phi = \sigma_2 \phi^* \sigma_2$ and
\bea
{\cal V}_\Delta &=& - \mu_3 ^2 [ Tr( \Delta_L \Delta_L ^\dagger) +
   Tr( \Delta_R \Delta_R ^\dagger)] + \rho_1 [ Tr( \Delta_L \Delta_L
^\dagger )^2 + Tr( \Delta_R \Delta_R ^\dagger )^2] + \nonumber \\ 
&&  \rho_2 [Tr( \Delta_L \Delta_L) Tr( \Delta_L ^\dagger \Delta_L
^\dagger) +       Tr( \Delta_R \Delta_R) Tr( \Delta_R ^\dagger
\Delta_R ^\dagger)] + \nonumber \\
&&\rho_3 Tr( \Delta_L \Delta_L ^\dagger) Tr(\Delta_R ^\dagger \Delta_R)
\eea
\newpage
\bea
{\cal V}_{\phi \Delta} &=&  \alpha_1 Tr( \phi^\dagger \phi) [ Tr(
\Delta_L \Delta_L ^\dagger) + Tr( \Delta_R \Delta_R ^\dagger)]
\nonumber \\
&&+ \alpha_2 [ Tr( \tilde \phi ^\dagger \phi)+ Tr( \tilde \phi
\phi ^\dagger) ] [ Tr( \Delta_L \Delta_L ^\dagger) + Tr( \Delta_R
^\dagger \Delta_R) ]
\nonumber \\
&& + \alpha_3 [ Tr (\phi \phi ^\dagger \Delta_L \Delta_L ^\dagger)
+ Tr (\phi ^\dagger \phi \Delta_R \Delta_R ^\dagger)] 
\nonumber \\
&& + \alpha_4 [ Tr (\tilde \phi \tilde \phi ^\dagger \Delta_L
\Delta_L ^\dagger) + Tr (\tilde \phi ^\dagger \tilde \phi
\Delta_R \Delta_R ^\dagger)]
\eea

The discrete LR symmetry ensures that all the couplings are real
and that the potential is CP conserving.  The scalar potential
may be simplified by imposing some more symmetry.  Thus, requiring
a $ \phi \rightarrow i\phi $ symmetry, the $\mu_2 ^2$  as well as
the $\lambda_4$ and $\alpha_2$ terms can be eliminated.
In the above expression for the scalar
potential, quartic terms of the form $\phi \Delta_{R} \phi
^\dagger \Delta_L ^\dagger + \phi ^\dagger \Delta_{L} \phi
\Delta_R ^\dagger$ have been excluded.   
In the literature these terms are usually dropped by appealing 
to some suitable symmetry \cite{rsymm}.

The  conditions  following from the minimization of the potential
are:
\beqn
k ^{'} [\lambda_1 (k^2 + k^{'2}) + 2(2 \lambda_2 + \lambda_3) k^2 
 + (\alpha_1 + \alpha_3) (v_L ^2 + v_R ^2) /2 - \mu_1 ^2]= 0 
\label{kpmin}
\eeqn
\beqn
k [\lambda_1 (k^2 + k^{'2}) + 2(2 \lambda_2 + \lambda_3) k^{'2} 
 + ( \alpha_1 + \alpha_4) (v_L ^2 + v_R ^2) /2 - \mu_1 ^2]= 0 
\label{kmin}
\eeqn
\beqn
v_L [ \ \rho_1 v_L ^2 + \rho_3 v_R ^2 /2 + \alpha_1 (k^2 + k^{'2} )/2
   + ( \alpha_3 k^{'2} + \alpha_4 k ^2)/2 - \mu_3 ^2 ] = 0 
\label{vlmin}
\eeqn
\beqn
v_R [ \ \rho_1 v_R ^2 + \rho_3 v_L ^2 /2 + \alpha_1 (k^2 + k^{'2} )/2
   + ( \alpha_3 k^{'2} + \alpha_4 k ^2) /2 - \mu_3 ^2 ] = 0 
\label{vrmin}
\eeqn
An examination of the charged gauge boson mass matrix (not
presented here) shows that the mixing between $W ^\pm_L$ and 
$W ^\pm_R$ is proportional to the product $k
k^{'}$. The tight limits on any right-handed admixture in the
observed weak interactions constrain this product
to be very near zero \cite{LRMIX} and in the remaining analysis we
take $k ^{'} =0$, in consonance with eq. (\ref{kpmin}). 
From eq. (\ref{vrmin}), since $v_R \neq 0$ to break $SU(2)_R$, one 
gets in this limit:
\beqn
\mu_3 ^2 = \rho_1 v_R ^2 + {{\alpha_1 + \alpha_4} \over 2} k ^2
+  \rho_3 v_L ^2 /2
\eeqn
Using this relation and $k^{'} =0$ in eq. (\ref{vlmin})  
one obtains:
\beqn
(2 \rho_1 - \rho_3) v_L (v_R ^2 - v_L ^2) =0 
\eeqn
This relation can be satisfied by choosing either $v_L = 0 $ or
$2 \rho_1 = \rho_3 $. If $v_L$ is non-vanishing then a global
symmetry of the theory -- identified with lepton number if
$\Delta_L$ has majorana couplings to leptons -- is spontaneously
broken resulting in a massless (goldstone) mode. Such a `triplet
majoron' is ruled out by the $Z$-decay data because the latter can
decay to a pair of such massless states with full strength, 
enhancing its invisible
decay width beyond the very stringent experimental constraints.
Therefore we choose the other alternative, namely, $v_L =0$
\cite{maj}.  The minimization conditions now become:
\beqn
\mu_1 ^2 = \lambda_1 k^2 + {{\alpha_1 + \alpha_4} \over 2} v_R
^2,\;\;\;
\mu_3 ^2 = \rho_1 v_R ^2 + {{ \alpha_1 + \alpha_4} \over 2} k ^2
\eeqn
along with $v_L,\; k^{'} =0$.  Utilizing the above relations,
the mass matrices for neutral and charged scalars in the
$\phi_1,\;\phi_2,\; \Delta_L,\; \Delta_R$ basis can be simplified.
Thus for real parts of the neutral scalars (CP even scalars):
\beqn
{\cal M}_{0r}^2 = \pmatrix{4 \lambda_1 k^2 & 0 & 0 & ( \alpha_1 +
\alpha_4) k v_R \cr
0 & 4(\lambda_3 + 2 \lambda_2) k^2 + (\alpha_3 - \alpha_4) v_R ^2
& 0 & 0 \cr 0 & 0 &( \rho_3 - 2 \rho_1) v_R ^2 & 0 \cr (
\alpha_{1} + \alpha_4 ) k v_{R} & 0 & 0 & 4 \rho_{1} v_{R} ^{2} }
\label{mr}
\eeqn
and for the imaginary parts (CP odd pseudoscalars):
\beqn
{\cal M}_{0i}^2 = \pmatrix{0 & 0 & 0 & 0 \cr
0 & 4(\lambda_3 - 2 \lambda_2) k^2 +( \alpha_3 - \alpha_4 ) v_R
^2 & 0 & 0 \cr 0 & 0 &( \rho_3 - 2 \rho_1) v_R ^2 & 0 \cr 0 & 0 &
0 & 0}
\label{mi}
\eeqn
For the singly-charged scalars one has:
\beqn
{\cal M}_{\pm}^2 = \pmatrix{( \alpha_3 - \alpha_4 )v_R ^2 /2 & 0 & 0 &
( \alpha_3 - \alpha_4) k v_R/2 \sqrt{2} \cr 0 & 0 & 0 & 0 \cr 0 & 0 &(
\rho_3 - 2 \rho_1) v_R ^2 /2 +
( \alpha_3 - \alpha_4 ) k^2 /4& 0 \cr ( \alpha_3 - \alpha_4 )k v_R /2 
\sqrt{2}& 0 & 0 & ( \alpha_3 - \alpha_4 ) k^2 /4} 
\label{m+}
\eeqn
while for the doubly-charged scalar mass matrix:
\beqn 
{\cal M}_{\pm \pm}^2 = \pmatrix{( \alpha_3 - \alpha_4 ) k ^2 /2 + 
(\rho_3 - 2 \rho_1) v_R ^2/2& 0   \cr 0 & 2 \rho_2 v_R ^2 
+ ( \alpha_3 - \alpha_4 )k ^{2} /2  } 
\label{m++}
\eeqn
As expected, there are two massless states each in ${\cal
M}_{0i} ^2$ and ${\cal M}_{\pm}^2$, corresponding to the
longitudinal modes of the gauge bosons $Z$, $Z^{'}$, $W_L$, and
$W_R$. From ${\cal M}_{0r} ^2 $, it is seen that $\phi^{0r} _2$
and $\delta_L ^{0r}$ are mass eigenstates.  The two other
eigenstates of this matrix are superpositions of $\phi_1^{0r}$
and $\delta_R^{0r}$  with a mixing angle of the order of $k \over
v_R$. The lighter eigenstate, with a mass of the order of $k
~(\sim m_W)$, is the analogue of the SM higgs boson.  Unless some
combinations of couplings are small (see later), the mass of all 
the other scalars
(including the singly-charged, doubly-charged and pseudoscalars)
are controlled by $v_R$.

The particular feature of the mass matrices in eqs. (\ref{mr} --
\ref{m++}) that we wish to stress in this work is that, in the chosen
$v_L = 0,\;\;k^{'}=0$ limit, all the scalars originating from
$\Delta_L$ are eigenstates of the corresponding mass matrices with the
contributions proportional to $v_R^2$ in the eigenvalues multiplied by
the factor $(\rho_3 - 2\rho_1)$.  If this last factor is small then
these states will be light.  Now, $(\rho_3 - 2\rho_1)$ will be exactly
vanishing if the gauge group is embedded in a simple grand unifying
group -- {\em e.g.,} $SO(10)$ Grand Unified Theory (GUT) -- so long as
{\em that} symmetry is unbroken since in the GUT $\Delta_L$ and
$\Delta_R$ are members of the same irreducible representation of the
symmetry group -- {\bf 126} of $SO(10)$, for example. The deviation of
the factor from zero can thus be considered as a result of the GUT
symmetry breaking.  We have examined the renormalisation group
evolution of these $\rho$ couplings from the GUT scale to the TeV
scale.  When writing down the evolution equations, for simplicity, we
have retained only the contributions from the gauge and $\rho$-type
quartic couplings. We find that for a range of values of the paramters
at the GUT scale, the magnitude of the combination $(\rho_3 -
2\rho_1)$ is in the appropriate ballpark \cite{rg_eqns}. Without
confining ourselves to any particular GUT, we examine the
phenomenology of the model assuming that the scalars originating from
$\Delta_L$ are not beyond the reach of the present and future
colliders.

A tree level relation among the masses of the scalars 
from the left-handed triplet is apparent from the matrices in
eqs. (\ref{mr} -- \ref{m++}), {\em viz.}:
\beqn
2 m^2 _{++} = 4 m^2 _{+} - m ^2 _{0}
\label{mreln}
\eeqn
Here $m_{++},\; m_{+}\;\; {\rm and}\;\; m_{0}$ are respectively
the masses of the doubly-charged, singly-charged, and the neutral
scalars.  In our subsequent analysis we vary the masses of the
scalars over a phenomenologically interesting range, consistent
with eq.  (\ref{mreln}), without delving  into the details of the
parameters of the scalar potential.

Here it might be mentioned that in our subsequent analysis the mass
ordering $m^2 _{++} > m^2 _{+} > m ^2 _{0}$ has been chosen. The
opposite hierarchy of the scalar masses, {\em viz.}, $m^2 _{0} > m ^2
_{+} > m ^2 _{++}$, can also be consistent with eq. (\ref{mreln}). But
it implies $(\alpha _3 - \alpha _4 ) < 0$ (see eqs. (\ref{m+} --
\ref{m++})) which in turn makes one of the charged scalar mass-squared
negative (see eq. (\ref{m+})).  Therefore, we do not conisder this
alternative.

Next we list the gauge boson - higgs boson interactions, relevant
for our investigation, in the convention where all the momenta
are incoming and each rule is to be multiplied by $i (p_1 -
p_2)^\mu \cos \xi ^0$ ($\xi ^0$ is the mixing angle in the $Z - Z
^{'}$ sector \cite{XI0}.  $p_1$ is the momentum of the first scalar boson
and $p_2$ is that of the second.):

\[
\delta^{0i} \delta^{0r} Z :  \frac{-i g}{\cos \theta_W},\;\;
\delta^{+} \delta^{-} Z: \frac{-g \sin^2 \theta_W}{\cos
\theta_W},\;\;
\delta^{+ +} \delta^{- -} Z: \frac{g \cos 2 \theta_W}{\cos
\theta_W}
\]

Finally, we note the majorana interaction of the triplet scalars
$\Delta $ with the leptons:
\begin{equation}
{\cal L} = i h \Psi^T_i C \tau_2 \Delta_i \Psi_i + h.c.;\;\; \Psi_i \equiv
\pmatrix{\nu \cr l}_i;\;\;i =  L,R 
\label{hcoupl}
\end{equation}
Here $C$ is the dirac charge conjugation matrix and $\tau_2$ the usual
$SU(2)$ generator. It turns out to be important for our later analysis
that the neutral member of the triplet couples only to neutrinos. In
this work, we assume the above interaction to be diagonal and
proportional to the identity in flavor space. This is the simplest
choice but certainly not unique. Off-diagonal entries will drive
lepton flavor violation and are constrained from processes like $\mu
\rightarrow e \gamma$, $\mu \rightarrow e e e$, $\tau \rightarrow \mu
\gamma$, etc. In view of the tight experimental bounds on these
\cite{PDG}, we take the liberty to drop the off-diagonal couplings.
Even keeping the couplings flavor diagonal, one might admit
non-universality. In the standard Yukawa sector, the couplings are
proportional to the fermion mass which is not the case here.  Rather,
we stick to the simplest choice of universal, diagonal couplings
\cite{r2}. We will illuminate this in some detail in the next section.

\section{MASS BOUND ON $\delta^0$ FROM LEP-II AND  PROSPECTS AT NLC}

As discussed in the previous section, we are interested in the $v_L =
0$ scenario. In this limit, $\delta^0$ does not have any coupling to a
pair of $Z$ bosons and cannot be searched for {\em via} a channel akin
to the Bjorken process for the SM higgs $H$ ({\em viz.} $e^+e^-
\rightarrow Z(Z^*) \rightarrow Z^* (Z) H$) at an electron-positron
collider. Instead, one must look for the production of $\delta^{0r}$
in association with a $\delta^{0i}$.  It is seen from eqs. (\ref{mr} -
\ref{m++}) that $\delta^{0r}$ and $\delta^{0i}$ are degenerate and are
the lightest among the members of the left-handed triplet.
Consequently, they will decay to a pair of neutrinos with a 100\%
branching ratio.  Thus, once produced in $e^+ e^-$ collision, they
will result in an invisible final state\footnote{The degeneracy of
$\delta^{0r}$ and $\delta^{0i}$ is a consequence of the $\phi
\rightarrow i \phi$ symmetry that has been imposed, ruling out certain
terms in the potential. In the absence of such a symmetry the
degeneracy will be removed. In such an event, the heavier of the two
will have an additional three body decay mode going to the lighter one
and a fermion-anti-fermion pair via a $Z^*$ exchange. We have checked
that for reasonable choices of the coupling $h$ (see
eq. (\ref{hcoupl})) this three body mode has a branching ratio of less
than 1\%.}.  Therefore, we examine the production of
$\delta^{0r}\;\delta^{0i}$ pairs accompanied by a photon which gives
rise to a single photon and missing energy signal. We have calculated
the $e^+ e^- \rightarrow \gamma \delta^{0r} \delta^{0i}$ cross-section
at a center of mass energy of 182.7 GeV corresponding to LEP-II.  The
main background for this signal comes from the SM process $ e^{+}e^{-}
\rightarrow \nu \bar{\nu} \gamma$. As the photon can originate only
from one of the initial electrons, a major part of the background will
be a peak in the distribution at the photon energy around 90 GeV. This
corresponds to $Z$ production exactly (or almost) on-shell. The
distribution also shows the usual {\em bremsstrahlung} peak for the
photon energy tending to zero. As regards the signal, since $m_0 <
m_Z/2$ is disfavored by the constraint from the $Z$ invisible width, a
bump in the photon energy is absent here.  Thus if this energy is
restricted to be inside a window that excludes the `radiative return
to Z' peak, the background cross-section is reduced significantly
without affecting the signal very much.  However, after imposing such
cuts we find it hardly possible to get a signal with $5 \sigma$
significance at LEP-II. We will return to this issue later in the
context of the Next Linear Collider (NLC) which will have higher
center of mass energy and more luminosity than LEP.  As discovery of
$\delta^0$ at LEP-II seems unlikely, it is of interest to look for the
bound on $m_0$ that can be obtained from the data.

To obtain the prediction of the LRM for this process at LEP-II, in
addition to the standard $Z$-exchange, the contribution from a
t-channel $\delta^{+}$ exchange diagram involveing the $\delta^+ e^+
\nu_e$ vertex has to be included.  Before proceeding further, let us
spend a few words on the upper bounds of these flavor diagonal
$\Delta$-lepton-lepton type couplings of majorana nature
\cite{Gunion,Raidal}.  The $\delta^+ e^+ \nu_e$ coupling is
constrained from the Bhabha ({\em viz.} $e^+ e^- \rightarrow e^+ e^-$)
scattering cross-section as follows.  In the LRM there is an extra
diagram contributing to Bhabha scattering {\em via} $\delta ^{++}$
exchange which is quadratic in this coupling.  The upper bound on it
increases with $m_{++}$ and, in turn (by virtue of the mass relation,
eq. (\ref{mreln})), with $m_+$.  This yields an upper limit on the
coupling of around 0.4 when $m_{+}$ is 200 GeV which increases to
about 4 when the $\delta ^{+}$ mass is of the order of 1 TeV.  The
upper bound for the diagonal $\mu \mu$ coupling can be as high as 10
derived from the measured $(g - 2)$ of the muon.  We found that in all
cases, the $\delta^{+}$ exchange t-channel diagram only makes a very
small contribution to the signal and does not affect our results
significantly. As already noted, we exclude flavor non-diagonal
couplings like $\delta^+ e^+ \nu_\mu$.

In Fig. 1 we present the number of $e^+ e^-\rightarrow
\gamma +~E\!\!\!/$ events in the LRM as a function of $m_0$. As
mentioned earlier, the LEP collaborations  have
searched for the single photon and missing energy final state at
LEP-II \cite{OPAL,OTHERLEP}.  We  use the OPAL results as they
are most conveniently adapted to our discussion.  The following
cuts have been used in line with ref.
\cite {OPAL}.
\begin{center}
$x^{\gamma}_{T} (\equiv E^{\gamma}_{T}/E_{beam})> 0.05,  \;\; 
165^0 > \theta_ {\gamma} > 15^0 $
\end{center}
The OPAL collaboration has observed $191$ $ \gamma + E\!\!\!/$
events in $e^{+} e^{-}$ collision at $\sqrt{s}$ = 182.7 GeV. The
SM prediction for this process is equal to $201.3 \pm 0.7$ events
at the same center of mass energy \cite{OPAL}.  We use these
numbers to set bounds on $m_0$ by requiring that the contribution
from the LRM must not exceed the difference between the 95\% C.L.
{\em upper} limit of the measured value and the 95\% C.L.  {\em
lower} limit of the SM prediction. This limit is indicated by the
dashed horizontal line in Fig. 1.  The lower bound on $m_0$
corresponds to the point of intersection of the cross-section
curve with this line and is seen to be equal to about 55.4 GeV.

Now we turn to investigate the prospect of discovery of the
$\delta^0$ at the Next Linear Collider with a center of mass
energy 500 GeV.  As earlier, we consider the $\gamma + E\!\!\!/$
signal and use the following cuts to enhance the signal over
background.
\begin{center}
$0.5 > x^{\gamma}_{T} > 0.05,  \;\; 
175^0 > \theta_ {\gamma} > 5^0, \;\;
$missing mass $>$ 100  GeV.
\end{center}
The matrix element squared for the background $e^+ e^-
\rightarrow \gamma \nu \bar{\nu}$ process for this analysis and
for the SM backgrounds in the following sections have been
calculated using the package MADGRAPH \cite{MAD} and the HELAS
\cite{HELAS} subroutines.  The cut on the photon energy and  on
the missing mass (invariant mass of the system recoiling against
the photon) helps to remove the events coming from the production
of an on-shell $Z$ and its subsequent decay to neutrinos, thus
reducing the background to a large extent. But the signal itself
is rather small at the NLC and, as evident from Fig. 2 where the
significance has been shown as a function of $m_0$, a $5 \sigma$
effect is possible upto about 110 GeV (and with $3 \sigma$ upto
150 GeV) essentially due to the higher luminosity.  For
a neutral scalar mass larger than 200 GeV, the number of signal
events falls rapidly so that the significance goes below 1.
Therefore, it will be possible to exclude  $m_0$ upto 160 -- 165
GeV at $95\%$ C.L. if no excess of $\gamma + E\!\!\!/$ signal is
seen at the NLC.

\section{LOOKING FOR THE $\delta^+$ AT TEVATRON AND LHC} 

Now we turn to the pair-production and detection possibilities of
the singly-charged scalar $\delta^+$ at hadron colliders.  Since
$v_L = 0$ has been chosen, $\delta^+W^-Z$ couplings are absent
and $\delta^\pm$ will decay only to a lepton and its neutrino. As
already mentioned, we assume that the $\delta^\pm$ couples to the
three lepton generations with equal strength and base this
discussion on the supposition that the leptonic decay modes are
dominant; {\em i.e.,} the $\delta ^\pm$ will decay to a lepton
and its anti-neutrino, branching ratio for each family being
$33.33\%$.  Thus, the pair produced charged scalars will give
rise to two leptons plus missing energy in the final state.

Before proceeding with the analysis, it is important to consider
what other decay modes might be possible for the $\delta^\pm$.
Since the couplings of the scalars are chosen diagonal in lepton
flavor, decays of $\delta^\pm$ to a charged lepton and a neutrino
of a different flavor are ruled out.  An important allowed mode
will be {\em via} the  $\delta ^{+} W_{L}^{-} \delta^{0}$
coupling.  This decay may be a two-body or a three-body one
depending on the kinematics.  Leptonic decay of the $W$ (on-shell
or off-shell) will result in the same  lepton plus missing energy
mode.  But this channel is suppressed by phase space.  Therefore
the assumption that the $\delta ^\pm$ decays to 3 families of
charged leptons and their neutrinos with $100 \%$ branching ratio
($b^+$ = 1, where $b^+$ is the total leptonic branching ratio),
may be an overestimate to some extent for higher $m_{+}$ but not
unrealistic. Some of these issues are discussed in greater detail
in Ref. \cite{Gunion}. (For comparison, we make brief remarks
about the results which obtain in the situation where the total
leptonic branching ratio is reduced to half this value.)

The SM background to the process is greatly reduced by
considering the case when the two final state leptons are of
different flavor. Since the detection of $\tau$ at a hadron
collider is not as efficient as that of the other leptons, our
stress is on the $e ^\pm \mu ^\mp p_T\!\!\!\!\!/$ ~signatures.
For these cases, there are no contributions to the background
from on-shell and off-shell $Z$ and photon Drell-Yan processes.
Further, there is no background from $l^+ l^- \gamma$ production
where the photon evades detection giving rise to missing $p_T$.
In the analysis presented below a parton level Monte-Carlo generator
using CTEQ-3M set of parton distributions \cite{PDFLIB} has been used.
  
The potential source of background for the above signal is from
on-shell and off-shell $W$ pair production.  Another class of
backgrounds arise from fake events where a quark jet is
misidentified as an electron or muon. Missing energy in such
events comes from two sources; 1) the mismeasurement of the jet
energy, and 2) from a gluon or photon which is radiated from a
quark and evades detection. The typical misidentification
probablity of a jet as an electron is around 2\%. The
contribution from the $qqg$ background (where the gluon is
outside the rapidity coverage and {\em both} quarks fake leptons)
is found to be about 0.8 {\em fb} to be compared with the
genuine $e \mu \, p_{T}\!\!\!\!\!/$ ~background of the order of
0.1 {\em pb}. Therefore, in the following analysis 
fake backgrounds are not considered any further.

Another source of $e \mu$ background would be from the decay of Drell
- Yan $\tau$ pairs. These $\tau$ pairs are back-to-back in the $p_T$
plane and are sufficiently energetic such that the electron or muons
coming from their decay are also moving in the same direction. So the
$e$ and $\mu$ are also back-to-back in the $p_T$ plane.  We applied a
$\Delta \phi _{e \mu} < 160 ^o$ cut which hardly reduces the signal
but removes this background completely.
 
We now turn to the detection possibilities. The kinematic
distribution of the signal and background events are similar. 
We present in Fig. 3 the missing $p_T$
distribution of the signal and the background at the Tevatron.
For the signal, a representative value of $m_+$ 
equal to 100 GeV has been chosen. One can see from Fig. 3
that for lower values of $p_T^{mis}$, the background is an order
of magnitude larger than the signal but it falls rapidly and
becomes less than the signal for $p_T^{mis} >$  100 GeV.
The $p_T$ spectrum of the leptons for the signal and the
background also show similar characteristics and are not
presented.

The following cuts are imposed in line with the Tevatron detectors:
\begin{center}
$p_T ^l > 35 ~{\rm GeV},\;\;
| \eta ^l | < 3,\;\;
\Delta R_{e \mu} > 0.5,\;\;
p_T ^{missing} > 80 ~{\rm GeV} $
\end{center}
The cut on the missing $p_T$ is very effective in reducing the
background. Still this
does not give enough boost to the signal-to-background ratio with
the present collected luminosity. But with an increased
luminosity of 25 fb$^{-1}$ (at Tevatron II) the $5 \sigma$
discovery limit extends upto a charged scalar mass of 140
GeV. This is shown in Fig. 4.

For LHC, the cross-section and luminosity being higher,  
more stringent cuts can be employed. For example, using
\begin{center}  
$p_T ^l > 50 ~{\rm GeV},\;\; 
| \eta ^l | < 3,\;\;
\Delta R_{e \mu} > 0.5,\;\;
p_T ^{\rm missing} > 100 ~{\rm GeV} $
\end{center}
in Fig. 4 we plot the significance of the signal. It can be seen
that the $5 \sigma$ discovery limit of $m_+$
is 240 GeV for an integrated luminosity of 100 fb$^{-1}$.

In this discussion, the branching ratio of the $\delta^+$ decaying to
leptons has been chosen to be $100 \%$ ($b^+ = 1$).  We can be more
conservative and assume the value of $b^+$ to be equal to 0.5 in which
case the signal and hence also the significance will reduce to one
fourth of their values above.  It is straightforward to check from
Fig. 4 that for both the Tevatron and the LHC a signal with $5\sigma$
significance can no longer be achieved.  For LHC the $2.5\sigma$
discovery limit is $m_+$ = 190 GeV.  It may be worth mentioning that
in the above analysis we always demand that the number of signal
events be at least equal to 5.

\section{PRODUCTION AND DETECTION OF $\delta^{++}$ AT LHC} 

We now turn to the production and detection of the doubly-charged
scalars, the heaviest among the members of the left-handed
triplet\footnote {To our knowledge, the detection possibilities of the
$\delta^{++}$ of the LRM at hadron colliders has not been discussed in
detail in the literature\cite{Dprod}. However, the production at
colliders of doubly-charged scalars, in general, has been examined.}.
First consider the pair production of $\delta^{++}$.  One might expect
that this production cross-section is lower than that for the lighter
$\delta^{+}$.  But this kinematic suppression is more than compensated
by the fact that the $\delta^ {++} \delta ^{--}$ coupling to $\gamma$
(and Z) is twice (almost) that of the $\delta^ {+} \delta ^{-}$.  The
$\delta^{++}$ will dominantly decay \cite{width} to a pair of like
sign leptons, thus giving rise to a spectacular four lepton signal. As
in the case of $\delta^+$, here again there are some other decay modes
of $\delta^{++}$ which may compete with the like-sign di-lepton
channel but over a wide mass range the latter dominates \cite{Gunion}.
When presenting the results for $\delta^{++}$ production and decay we
set the $ll$ branching ratio to 100\% ($b^{++}$ = 1).  Results for any
other value of $b^{++}$, say $x$, can be readily obtained from these
since the number of signal events is simply reduced by an appropriate
factor, $x^2$, while the background is unaffected.
 
The main source of background for this process is from the hadronic
four lepton production. Considering the case when one scalar decays to
a pair of electrons $(e^- e^-)$ while the other decays to a pair of
muons $(\mu^+ \mu^+)$, one can easily see that this final state has no
SM background at all since the detector can identify \cite{CMS} the
particle charge\footnote {If all the leptons are of the same flavor
then the signal would be {\em e.g.,} $e^+ e^+ e^- e^-$. The SM
background for this particular final state is small but
non-zero. However, the requirement $M_{e ^- e ^-} = M_{e ^+ e ^+}$
will remove these $4e$ SM events.}

We apply the following set of acceptance cuts for the LHC:
\begin{center}
$p_T ^l > 20 ~{\rm GeV},\;\;
| \eta ^l | < 3.0,\;\; 
\Delta R_{l l} > 0.5  $
\end{center}
In Fig. 5 we plot the number of expected events as a function of
$m_{++}$. As there is no background for this signal, 5 events may be
considered as a benchmark for the discovery. From the figure we see
that one can go upto a $\delta ^{++}$ mass of 850 GeV with a proposed
LHC integrated luminosity of 100 fb$^{-1}$.  If $b^{++}$ equals 0.5,
{\em i.e.,} the leptonic modes account for only half the total width,
then $m_{++}$ upto about 640 GeV can be explored satisfying the `5
events' criterion.

One may be more conservative and ask about the significance of the
signal if charge identification of the leptons is not very efficient
\cite{CMS}. For this purpose, we also calculated the SM cross-section
for the process $p p \rightarrow e ^+ e^- \mu^+ \mu^-$ using a parton
level Monte-Carlo generator. We assume here the worst case scenario,
namely, that one cannot identify the lepton charges at all. Thus the
above background can mimic the signal. The main contribution to the
background originates from the on-shell production of a pair of $Z$s.
We also note that for signal events the invariant mass for the pair of
electrons must be equal to that of the muon pair.  Apart from the
kinematic cuts given above we impose further cuts, $ M_{e e} \simeq
M_{\mu \mu} > 100 ~{\rm GeV} $. This removes all events coming from
the decay of two on-shell $Z$s. The presence of the background cannot
reduce the prominence of the signal very much.  The number of
background events in the bins of invariant mass are orders of
magnitude less than those from the signal. At $M_{ee}$ = 150 GeV, the
number of background events is around .01 compared to a signal of $6
\times 10^3$.  The SM background falls to $10^{-6}$ at $M_{ee}$ = 600
GeV where the signal from LRM is nearly equal to 20.  This is mainly
due to the kinematic cut used to remove the events coming from the
decay of the $Z$s. For the remaining events, either from off-shell $Z$
or photon, it is highly unlikely that the invariant mass of the muon
pair will be the same as that of the electron pair. Thus,
non-observation of this spectacular $e e \mu \mu$ signal at the LHC
will help to exclude the $\delta^{++}_L$ mass upto about 650 GeV.

We have also estimated the prospect of a search for the
$\delta^{++}$ at the Tevatron. But with the lower center of mass
energy compared to the LHC and with a luminosity of 25 fb
$^{-1}$ ({\em i.e} at the upgraded Tevatron) one can reach
$m_{++}$ upto only 275 GeV if 5 events are set as a
benchmark for discovery (see Fig. 5).
\vskip 0.25in

\section{CONCLUSIONS} 
We examined the minimal version of the Left-Right Symmetric model,
specifically the phenomenology of the left-handed triplet scalars. In
general, all the scalars barring one can be heavy (of the scale 1 TeV
or more). But some choices of the scalar potential couplings, which
are motivated by GUT ideas, can keep the masses of the scalars from
the left-handed triplet in the scale of $m_W$.  The neutral one of
these scalars can only decay to neutrinos. So this decay is
invisible. We have put a new bound on the mass of such a neutral
scalar from the LEP-II measurement of the $e^+ e^- \rightarrow \gamma
+ E\!\!\!/$ cross-section and examined the reach of the NLC to explore
$m_0$.

We have also investigated the production and detection of the
singly-charged and doubly-charged scalars at the Tevatron and the
LHC and compared the signal with the SM background.  For the
doubly-charged scalar signal there are basically no SM background
events to compare with. The detection of such doubly-charged (as
well as singly-charged) scalars at the LHC look quite promising.
At the Tevatron with increased integrated luminosity, detection of
a singly-charged scalar is possible upto a mass of 140 GeV
while for the LHC the mass reach for $5\sigma$ discovery is
around 240 GeV.  The doubly-charged scalar can be probed at the
LHC upto a mass of about 600 GeV.

\rm

\vskip 30pt
%\newpage
{\large {\bf Acknowledgements}}
\normalsize\rm

This research is  supported by the Council of Scientific and
Industrial Research, India.  AR acknowledges partial support from
the Department of Science and Technology, India.

\newpage

\begin{center}
{\bf FIGURE CAPTION}
\end{center}

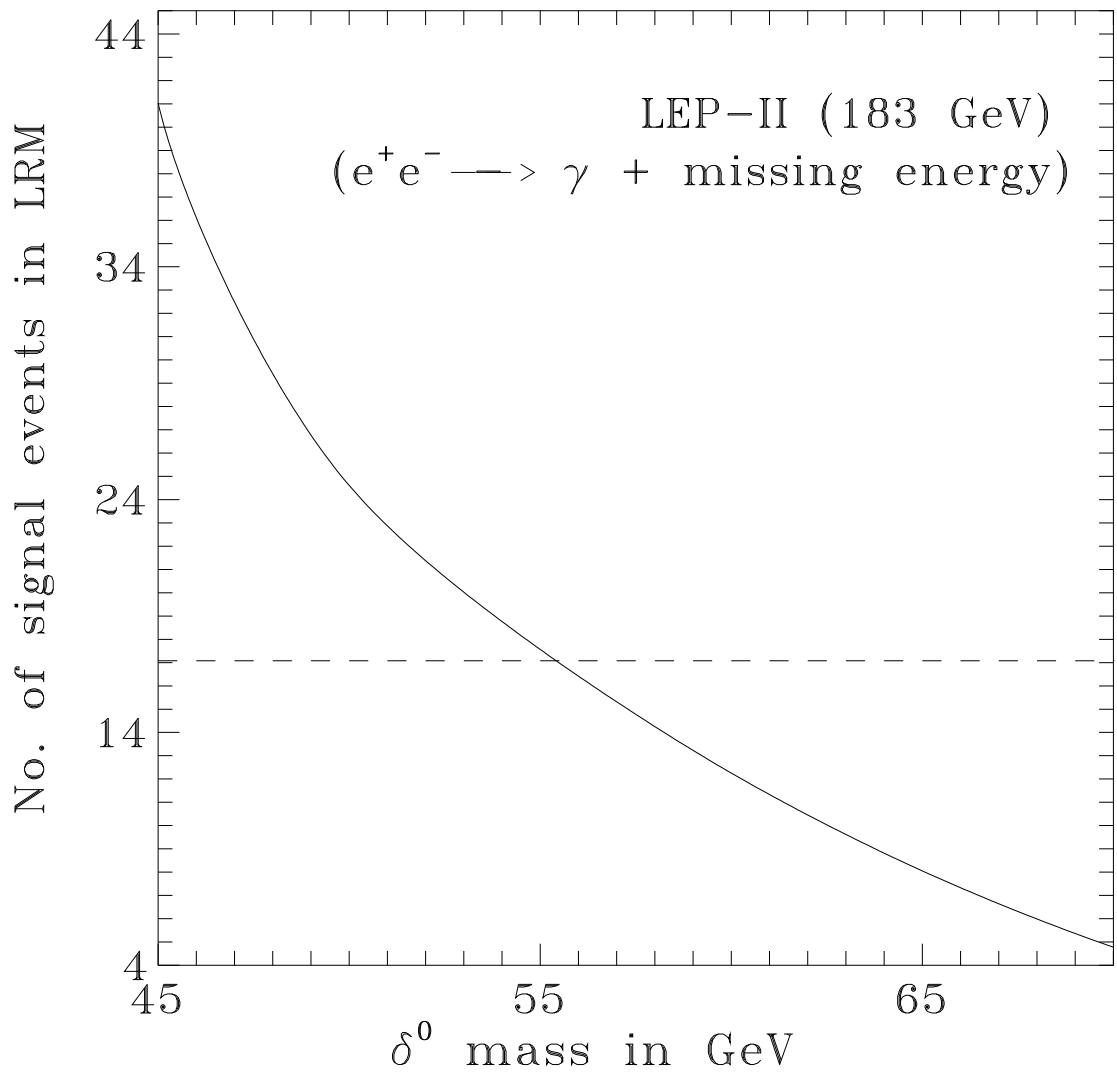
Fig. 1.  The number of $e^+ e^- \rightarrow \gamma + E\!\!\!/$ events
as a function of the $\delta^0$ mass in the Left-Right symmetric
model. The dashed horizontal line is the difference between the 95\%
C.L. upper limit of the observed number of events at LEP-II with
$\sqrt{s}$ = 182.7 GeV and the lower limit of the SM prediction at the
same C.L.  The lower limit on the $\delta^0$ mass is determined by the
point of intersection of the solid curve with the dashed line.

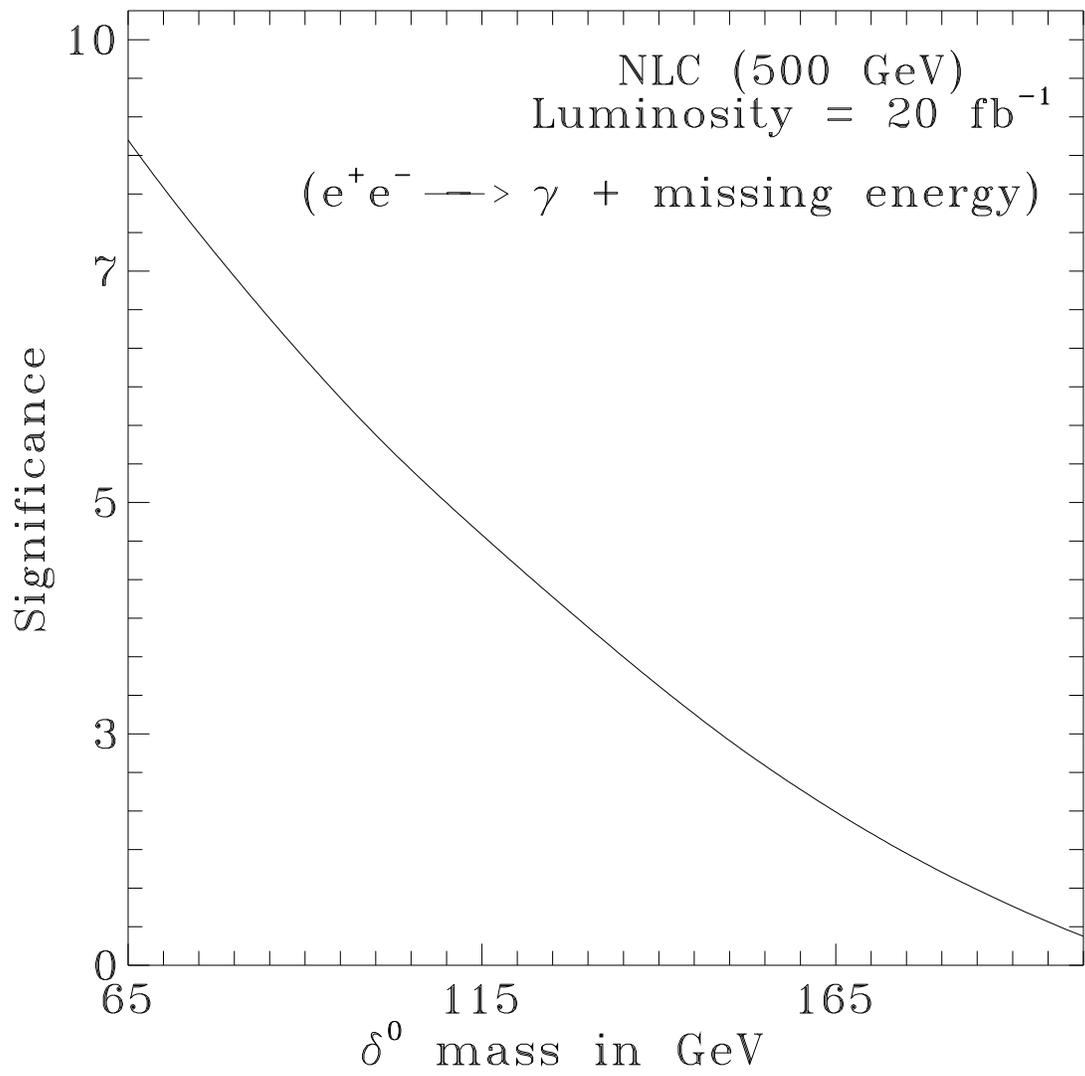
Fig. 2. Significance ($\equiv {\rm Signal}/\sqrt{\rm Background}$) for
the process $e^+ e^- \rightarrow \gamma + E\!\!\!/$ as a function of
$\delta^0$ mass in the Left-Right symmetric model at the NLC with 20
fb$^{-1}$ integrated luminosity and $\sqrt{s}$ = 500 GeV.

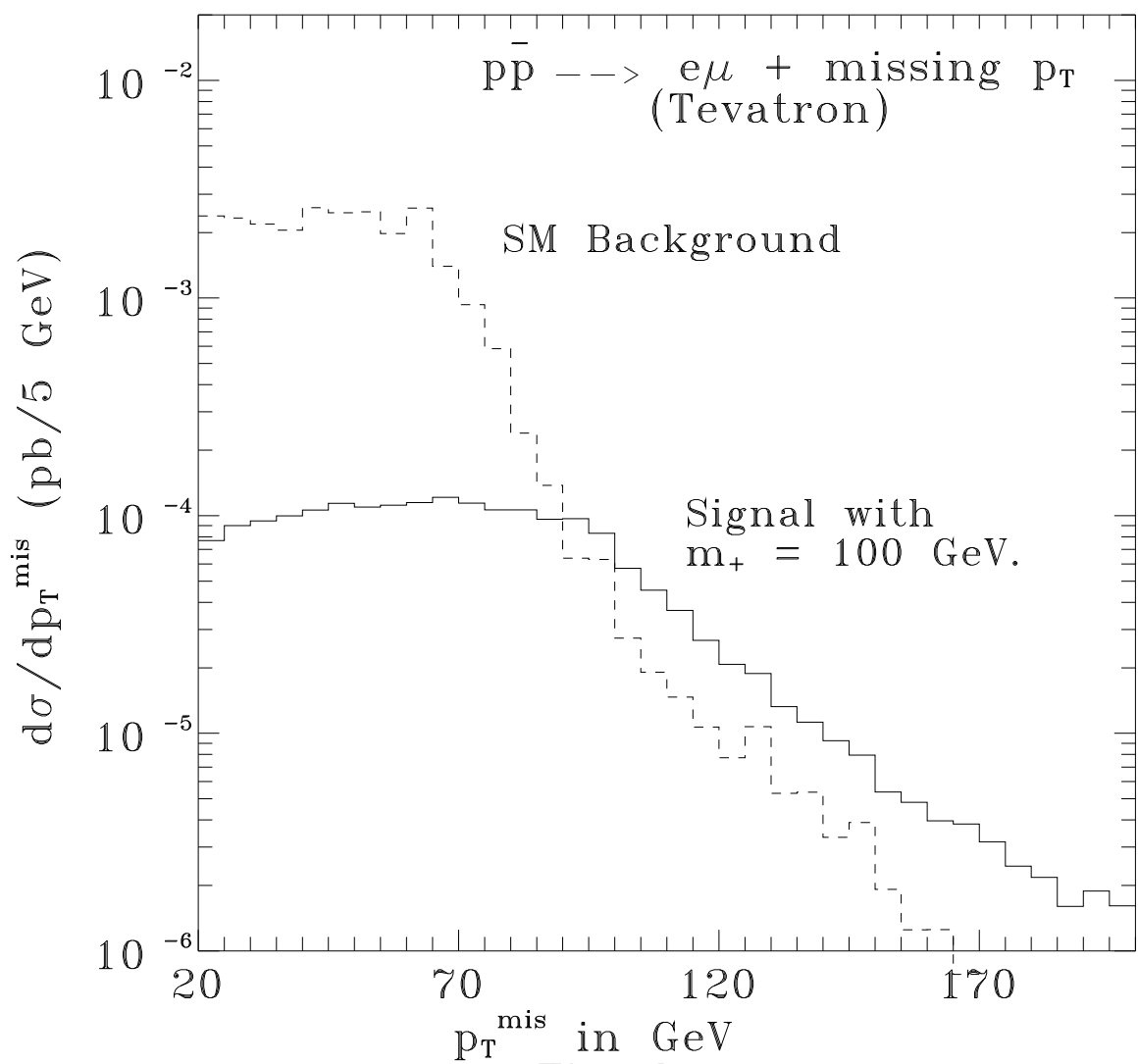
Fig. 3. Missing $p_T$ distribution for the process $p \bar{p}
\rightarrow e \mu + p_T\!\!\!\!\!/$ ~from the Left-Right symmetric
model signal (solid histogram) and the SM background (dashed
line) at the Tevatron.  For the signal, a representative value of
the charged scalar mass equal to 100 GeV is chosen.

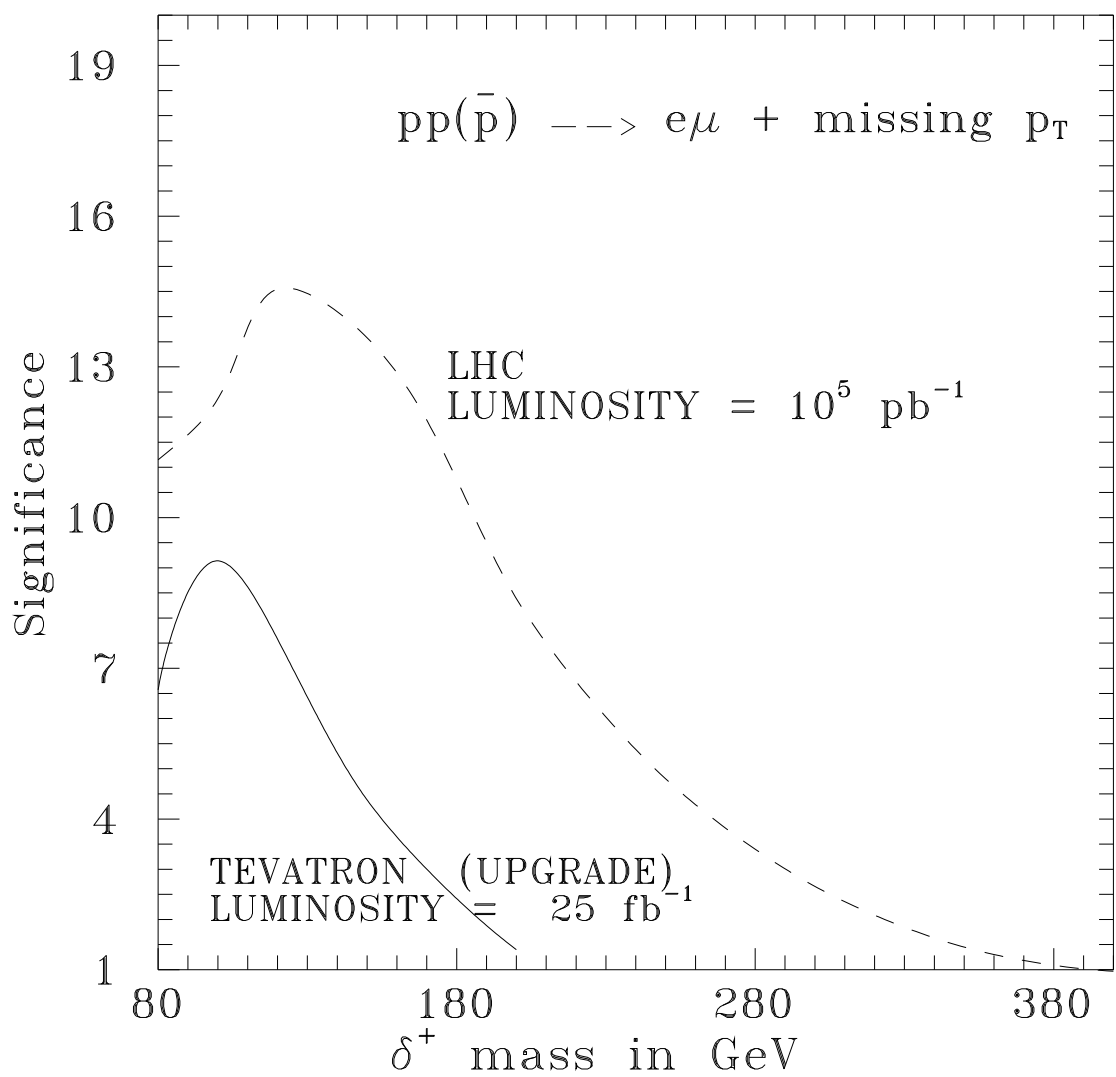
Fig. 4. Significance ($\equiv {\rm Signal}/\sqrt{\rm
Background}$) for the process $p \bar{p}$  or $p p \rightarrow 
e \mu + p_T\!\!\!\!\!/$ ~in the Left-Right symmetric model 
as a function of
$\delta^\pm$ mass for the Tevatron with an integrated luminosity
of 25 fb$^{-1}$ (solid line) and for the LHC ($\sqrt{s}$ = 14
TeV) with an integrated luminosity of 100 fb$^{-1}$ (dashed
line).

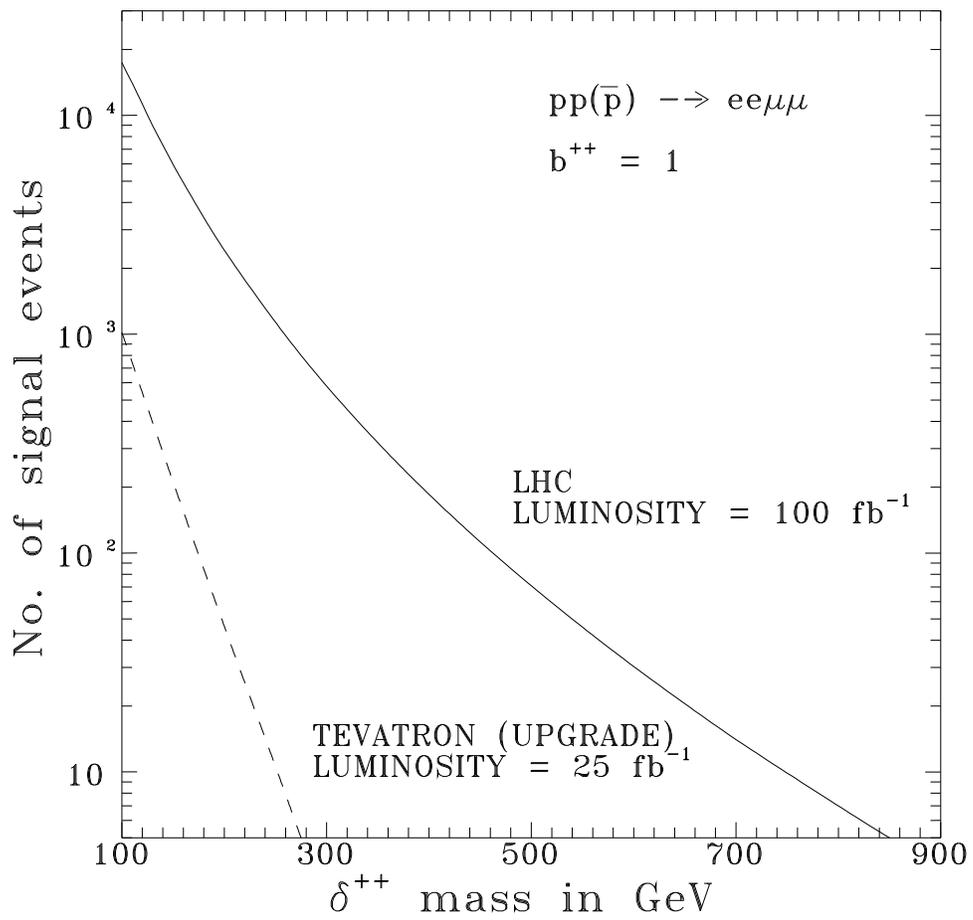
Fig 5. Number of  signal events  for $p p \rightarrow e^-  e^-
\mu^+ \mu^+ $ from $\delta^{++}$ production as a function of $ee$
or $\mu\mu$ invariant mass for the Tevatron with an integrated luminosity
of 25 fb$^{-1}$ (dashed line) and for the LHC ($\sqrt{s}$ = 14
TeV) with an integrated luminosity of 100 fb$^{-1}$ (solid
line).


\begin{thebibliography}{99}
\bibitem{maj} Notice that one can no longer generate 
majorana masses for the left-handed neutrinos. However, in the
usual see-saw formulation, this entry of the
mass matrix is often neglected anyway.
\bibitem{rg_eqns} For example, if we choose $\rho _1 = .05,\; \rho
_2 = .1, \;\rho _3 = .1$ at the GUT scale satisfying the $\rho_3 -
2\rho_1 = 0$ requirement, the magnitude of $\sqrt{\rho _3 - 2\rho
_1}$ becomes $0.05$ at the TeV scale, which is in the appropriate
range. On the other hand the couplings become greater than unity
at the TeV scale if they are assigned large values like $\rho _1
= .5,\; \rho _2 = 1,\; \rho _3 = 1$ at the GUT scale.
\bibitem{width}The width of the decay $\delta ^{++} \rightarrow l ^+ l
^+$ can be written as $\simeq h _{ll} ^2 m _{++} /25$ (see, for
example, the second paper in \cite{Dprod}).  If $h$ is of the order of
$0.1$ then this width is of the order of $0.2$ GeV for $m _{++} \sim$
500 GeV.
\bibitem{CMS}CMS, Technical Proposal, CERN/LHCC 94-38, LHCC/P1; 
The charge assignment is correct at 99\% C.L. upto a muon
momentum  of 7 TeV for full rapidity coverage.
\end{thebibliography}
\end{document}